%% file: MAIN5.tex
\RequirePackage{fix-cm}
\documentclass[smallcondensed]{svjour3}     % onecolumn (ditto)
\AtBeginDocument{ \paperwidth=\dimexpr 1in + \oddsidemargin + \textwidth + 1in + \oddsidemargin \relax \paperheight=\dimexpr 1in + \topmargin + \headheight + \headsep + \textheight + 1in + \topmargin \relax \usepackage[pass]{geometry} \relax }
\usepackage[T1]{fontenc} %together1
\usepackage{tgtermes, tgheros} %together1
\usepackage{amsfonts}
\usepackage{amsmath}

\usepackage{amsthm}
\usepackage{mathtools}
\usepackage{optidef}
\usepackage{stmaryrd}
\usepackage{graphicx, color}
\usepackage{hyperref}
\usepackage{natbib}
\usepackage{caption}
\usepackage{subcaption}
\usepackage{layout}
\usepackage{comment}
\usepackage{setspace}
\usepackage{lscape}
\usepackage{dcolumn}
\usepackage{multirow}
\usepackage{threeparttablex}
\usepackage{tabularx}
\usepackage{longtable}
\usepackage{afterpage}
\usepackage{commath}
\usepackage{comment}
\usepackage[table]{xcolor}
\usepackage{array, booktabs}

\newtheorem{assumption}{Assumption}

\DeclareMathOperator*{\plim}{plim}
\DeclareMathOperator*{\cov}{cov}
\journalname{ }
\def\makeheadbox{\relax}
\begin{document}
\title{
How do we measure trade elasticity for services?
}
\titlerunning{ }        % if too long for running head
\author{Satoshi Nakano \and Kazuhiko Nishimura}
\institute{
Satoshi Nakano \at Nihon Fukushi University, Tokai, 477-0031 \\
\email{nakano@n-fukushi.ac.jp}
\and
Kazuhiko Nishimura \at Chukyo University, Nagoya, 466-8666 \\
\email{nishimura@lets.chukyo-u.ac.jp}
}
\date{\today}

\maketitle
\begin{abstract}
This paper presents our attempt at identifying trade elasticities through variations in the exchange rate for possible applications to the case of services whose physical transactions are veiled in trade statistics.
The regression analysis used to estimate elasticity entails a situation where the explanatory variable is leaked into the error term through the latent supply equation, causing an endogeneity problem for which an instrumental variable cannot be found.
Our identification strategy is to utilize the normalizing condition, which enables the supply parameter to be identified, along with the reduced-form equation of the system of demand and supply equations.
We evaluate the performance of the method proposed by applying it to several different tangible goods whose benchmark trade elasticities are estimable by utilizing information on their physical transactions.
\end{abstract}

\clearpage
\section{Introduction}
\begin{figure}[t!]
\centering
\includegraphics[width=0.49\textwidth]{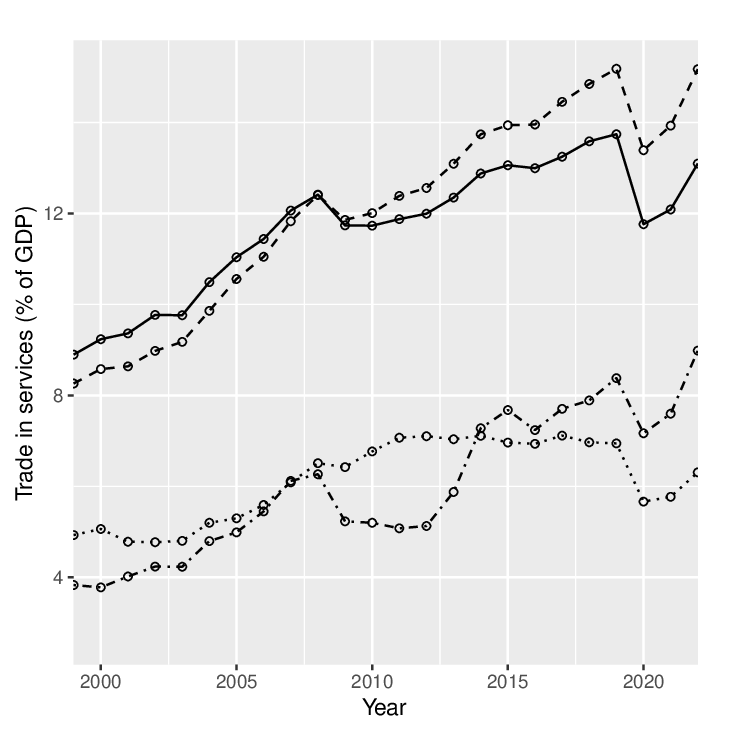}
\includegraphics[width=0.49\textwidth]{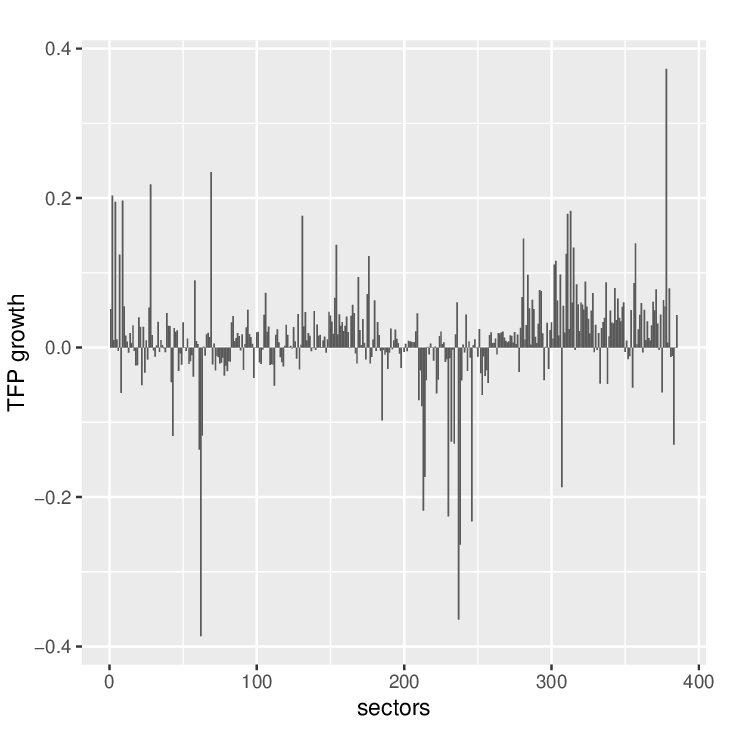}
\caption{
Left: Trade in services measured in percent share of GDP from 1999--2022.
The solid, dashed, dotted and dash-dotted lines correspond to the world, OECD, US, and Japan, respectively (source: The World Bank).
Right: Sectoral total factor productivity growth from 2000--2005 for
Japan, on the basis of the T\"{o}rnqvist index.
Service sectors (tertiary industries) are indexed between 283 and beyond (source: 2000--2005 linked input-output tables for Japan).
} \label{fig_intro}
\end{figure}

Trade in services is rapidly gaining presence in the world economy (see Figure \ref{fig_intro} left).
Furthermore, the productivity of service sectors is growing (see Figure \ref{fig_intro} right), making services less expensive in real terms. 
Under these circumstances, it is tempting to study the impact of increased productivity of service sectors on the structure of trade in services.
As we follow the demand-driven trade theory, the level of exportation depends on the import demand from various countries, while the import demand depends on the prices of the commodity from various countries on the soil of the importing country and, in particular, its elasticity of substitution among the commodity from different countries (i.e., trade elasticity, or Armington elasticity).\footnote{
Armington elasticity is the key parameter in international trade policy analysis under the general equilibrium modeling framework \citep{baj2020}
}
For trade in services, the latter has not been measured by means of
standard techniques, since a physical measure for services is lacking in trade statistics.

The trade elasticity for services, nevertheless, has been the subject of measurement in the literature \citep[e.g.,][]{oecd, christen, blank}.
Theory states that the supply under monopolistic competition together with the demand derived from the CES utility associates the markup ratio of the monopolistic supplier with the elasticity of substitution.
To project this theory onto international trade, an exporting country
as a whole must behave as a monopolist facing a downward-sloping demand schedule of the differentiated commodity consumed in the importing country.\footnote{
In other words, such an assumption is needed to measure trade elasticities from sector-level data, as in \citet{egger}.
}
Previous studies measure the trade elasticity of services by the markup ratios obtained from firm data, implicitly framing the elasticity measurement of substitution among intranational varieties.
Such elasticity is, of course, not what we are looking for.

For tangible goods, trade elasticity is usually measured by regressing import value shares onto the prices observed on the soil of the importing country.
The price in this context equals the product of 1) the price at
shipment from the exporter's border (i.e., the FOB price), 2) a factor
representing various transportation costs (or the CIF/FOB ratio), 3)
the ad valorem equivalent tariff rate, and 4) the rate to exchange the exporter's currency into the importer's currency.
This means that the regression coefficient (with which elasticity is measured) is common to all four factors 1) \textendash 4).
Because of this property, \citet{nber} and \citet{tariffbased} focus on identifying trade elasticities entirely through variations in tariffs.\footnote{
In these studies, however, country-specific elasticities are not of concern, whereas in this study, they are.
}
While the same strategy cannot be applied to services (because ad valorem tariffs are not levied on services), we may consider using the fourth factor: the exchange rate.

In that event, we still face omitted variable bias, as we are omitting
the primary factor, i.e., 1) the export price, from the explanatory variables of the regression.
The fundamental source of the problem is the latent supply function,
where the importer's demand (the response variable) is fed back into
the exporter's price, which is then included in the error term.
In this situation, the explanatory variable (i.e., the exchange rate) will leak into the error term, causing an endogeneity problem. An instrumental variable to fix the bias pertaining to such endogeneity, therefore, must be uncorrelated with the exchange rate leaked into the error term and, yet, must strongly be correlated with the explanatory variable, that is, the exchange rate.
The nonexistence of such objects leads us to exclude the instrumental variable approach from our work.

Our identification strategy is to utilize two endogeneity-free regression equations.
The first is the equilibrium equation, which combines the demand and supply functions.
This identifies a mix of demand and supply parameters.
The second is the price-normalizing condition in which the true share parameter agrees with the value share.
This condition is a cross-sectional regression equation that identifies the supply parameter without the concern of endogeneity.
The trade elasticity is thus estimated by the two equations, using observations on import value shares and exchange rates, in a panel setting.
We assess the performance of this approach with respect to other means
(i.e., Feenstra's method \citep{feenstra_aer94, HH, feenstra_restat18}
and extensions) to combat endogeneity without employing instrumental variables.

We evaluate the performance of the abovementioned alternatives by
applying them to several different tangible goods whose elasticities are estimated via the standard technique with instrumental variables.
As mentioned above, trade elasticity can be estimated by the prices where all four factors are monitorable for the case of tangible goods.
Nevertheless, the latent supply function will feed the importer's
demand into the exporter's price, but now it is observed as the FOB price and is part of the explanatory variables of the regression, not in the error term.
The error term thus represents a pure demand shock for the importer.
In this situation, the endogeneity caused by the latent supply function can be remedied by the exchange rate, which must be correlated with the explanatory variable of being its own component but uncorrelated with the error term of being independent of demand shocks.\footnote{\cite{cje02} \cite{nn25} employ the exchange rate as an instrument to combat endogeneity when estimating trade elasticities for tangible goods.}

The remainder of the paper proceeds as follows.
In the following section, we introduce the model of Armington
aggregation, deriving the regression equations for estimating the
trade elasticity, while specifically addressing concerns of reverse causality and omitted variable endogeneity.
We also explain our method of estimating trade elasticities from value shares and exchange rates.
In Section 3, we first estimate benchmark trade elasticities for
several tangible goods by means of a standard technique applicable to the case where prices are available.
We then proceed to assess the performance of alternative methods that utilize exchange rates instead of prices.
We finally apply alternative methods to estimate the trade elasticity for services.
Section 4 concludes the paper.
The replication data for this study are available at \citet{dataverse}.

\section{Analytical framework}
\subsection{Model \label{sec_model}}
Below are an Armington aggregator and the corresponding dual function for a kind of service $g$ (index suppressed) imported into a home country from $N$ different countries:
\begin{align}
Y = \left( \sum_{i=1}^N (\lambda_i)^{\frac{1}{\sigma}} (X_i)^{\frac{\sigma-1}{\sigma}} \right)^{\frac{\sigma}{\sigma-1}},
&&
Q = \left( \sum_{i=1}^N \lambda_i (P_i)^{1-\sigma} \right)^{\frac{1}{1-\sigma}}
\label{dual}
\end{align}
where $X_i$ denotes the virtual quantity (of service) imported from
the $i$th country, and $P_i$ denotes the corresponding virtual price in
terms of the home country's LCU (local currency unit).
Let us hereafter call $P_i$ the import price.
Similarly, $Y$ denotes the partial utility for aggregated imports, and $Q$ denotes the corresponding unit value.
For the parameters, $\sigma$ denotes the elasticity of substitution among imports from different countries (i.e., Armington elasticity), and $\lambda_i \geq 0$ denotes the preference parameter where $\sum_{i=1}^N \lambda_i =1$.
The duality asserts that $QY = \sum_{i=1}^N P_i X_i = \sum_{i=1}^N V_i$ in all cases.

Applying Shephard's lemma for the dual function yields the following:
\begin{align}
\frac{P_i}{Q} \frac{\partial Q}{\partial P_i} 
= \frac{P_i}{Q}\frac{X_i}{Y}  = \frac{V_i}{\sum_{i=1}^N V_i}
= S_i
= \lambda_i \left( \frac{P_i}{Q}\right)^{1-\sigma}
\label{foc}
\end{align}
where $S_{i}$ denotes the value share of imports from the $i$th country.
As we index the observations by $t=1,\cdots,T$, the above condition implies the following regression equation:
\begin{align}
\ln S_{it} &= \ln \lambda_i -\gamma\ln Q_{t} + \gamma \ln P_{it}  +  \varepsilon_{it} 
\label{regmain}
\end{align}
where $\gamma = 1-\sigma$ and where $\varepsilon_{it}$ denotes the disturbance term.\footnote{
The disturbance term consists of demand shocks, including myriad nontariff barriers.
\ref{appdx3} provides some analysis based on survey data for nontariff barriers to trade in services.
}
At a glance, we may consider estimating $\gamma$ via two-way fixed effects estimation, eliminating the country-specific as well as the time-specific effects, by double-demeaning on both sides of the equation as follows:
\begin{align}
\left[ \ln S_{it} \right] = \gamma \left[ \ln P_{it} \right] + \left[ \varepsilon_{it} \right]
\label{regdd}
\end{align}
where square brackets indicate double-demeaning.\footnote{
Specifically, for example, $\left[ \varepsilon_{it} \right] =\varepsilon_{it} - \frac{1}{T}\sum_{t=1}^T \varepsilon_{it} - \frac{1}{N}\sum_{i=1}^N \varepsilon_{it} + \frac{1}{NT}\sum_{i=1}^N\sum_{t=1}^T \varepsilon_{it}$.
}

Of course, regression (\ref{regdd}) is impossible since the explanatory variable $P_i$ is only partly available.
Specifically, $P_i=Z_i\pi_i$ is available for part of the exchange rate $Z_i$ but unavailable for part of the service's export price $\pi_i$ in terms of country $i$'s LCU.
Moreover, even if $\pi_i= V_i/X_i/Z_i$ were available, such as for any nonservice commodities whose $X_i$ values are available, regression (\ref{regdd}) would suffer from an endogeneity problem, since the export price $\pi_i$ must be affected by the import demand shock $\varepsilon_i$ through the supply function.
Below, we write the corresponding demand and supply equations:
\begin{align}
\ln S_{it} &= \ln \lambda_i  -\gamma \ln Q_t + \gamma \left( \ln Z_{it}  + \ln \pi_{it}  \right) +  \varepsilon_{it}  \label{regsimul1} \\
\ln \pi_{it} &= \tau + \omega \ln S_{it}  +  \delta_{it} 
\label{regsimul2}
\end{align}
The demand equation (\ref{regsimul1}) is merely a copy of Equation (\ref{regmain}).
The supply parameter $\omega$ denotes the elasticity of the export price $\pi_{it}$ with respect to the value share demanded $S_{it}$, and $\delta_{it}$ denotes the disturbance term concerning supply shocks.
For the sake of simplicity, the supply equation (\ref{regsimul2}) does not have a time- or country-specific term but rather a constant, $\tau$.
In addition, we believe that it is legitimate to make the following assumption.\footnote{Independence between demand and supply shocks is a common assumption \citep[made in, e.g., ][]{feenstra_aer94, feenstra_restat18}.
An obvious small market assumption of a single commodity against the entire currency market implies noncorrelation between demand/supply shocks and currency exchange rates.
A similar assumption is made in \citet{cje02} and \citet{nn25}.
}
\begin{assumption}\label{ortho}
The supply shocks $\delta_{it}$, demand shocks $\varepsilon_{it}$, and (growth rate of) exchange rates $\ln Z_{it}$ are mutually independent, i.e.,
\begin{align} 
\cov (\delta_{it}, \varepsilon_{it}) = \cov (\delta_{it}, \ln Z_{it}) = \cov (\ln Z_{it}, \varepsilon_{it})=0
\end{align}
\end{assumption}

Let us modify the demand equation further as follows:
\begin{align}
\ln S_{it} &= \ln \lambda_i  -\gamma \ln Q_t + \gamma \ln Z_{it}  + \left( \gamma \ln \pi_{it}  +  \varepsilon_{it} \right) 
\label{regwithz}
\end{align}
where the parentheses indicate the error term.
We can double-demean both sides of Equation (\ref{regwithz}) to yield the following equation:
\begin{align}
\left[ \ln S_{it} \right] = \gamma \left[ \ln Z_{it} \right] + \left[ \gamma \ln \pi_{it}  +  \varepsilon_{it} \right]
\label{regdd2}
\end{align}
Unlike Equation (\ref{regdd}), the explanatory variable for this equation is available.
However, an endogeneity problem lies in that the explanatory variable is likely to be correlated with the error term, as $Z_{it}$ enters $\pi_{it}$ via $S_{it}$, according to Equations (\ref{regsimul1}) and (\ref{regsimul2}). 
Moreover, there is no way to find an instrument to remedy this problem.
The explanatory variable $Z_{it}$ (exchange rate for the $i$th country) is considered exogenous (i.e., established outside of the market of the service concerned), and while it may be easy to find variables that are correlated with $Z_{it}$, we can never be certain that any one such variable is uncorrelated with the invisible $\pi_{it}$ that we suppose is contaminated with $Z_{it}$.

\subsection{Identification strategy}
Let us eliminate the latent export price $\pi_{it}$ from the demand and supply equations (\ref{regsimul1}) and (\ref{regsimul2}) as follows:
\begin{align}
\ln S_{it} = \frac{1}{1-\gamma \omega} \ln \lambda_i + \frac{\gamma}{1-\gamma \omega} (\tau - \ln Q_t) + \frac{\gamma}{1-\gamma \omega} \ln Z_{it} + \frac{\gamma \delta_{it} + \varepsilon_{it}}{1-\gamma \omega}
\label{eeq}
\end{align}
and call this reduced form the equilibrium equation.
If we know the true share parameter $\lambda_i$, this regression
equation could identify both $\gamma$ and $\omega$, where by
Assumption \ref{ortho}, without endogeneity problems.
While we cannot know the true share parameter, we know by virtue of equations (\ref{dual}) and (\ref{foc}) that they coincide with the value shares at the state where $P_i =1$ for $i=1,\cdots,N$.
Therefore, we let ourselves virtually normalize all import prices at some point in time $t=\theta$ as follows:
\begin{align}
\ln P_{i\theta} = \ln Z_{i\theta} + \ln \pi_{i\theta} = 0 &&
i=1, \cdots, N
\label{norma}
\end{align}
By virtue of equation (\ref{norma}), the demand and supply equations become as follows:
\begin{align}
\ln S_{i\theta} &= \ln \lambda_i  -\gamma \ln Q_{\theta} 
+  \varepsilon_{i\theta}   \label{regdemand}\\
- \ln Z_{i\theta} &= \tau + \omega \ln S_{i\theta}  +  \delta_{i\theta}  \label{regsupply}
\end{align}
Note that (\ref{regsupply}) is a regression equation with which the supply parameter $\omega$ can be identified.
Furthermore, Equation (\ref{regdemand}) indicates that regression equation (\ref{regsupply}) is endogeneity-free, in light of the following expansion:
\begin{align}
\cov \left( \ln S_{i\theta}, \delta_{i\theta} \right) 
&= \cov \left( \ln \lambda_i - \gamma \ln Q_{\theta} + \varepsilon_{i\theta}, \delta_{i\theta} \right) 
\\
&= \cov \left( \ln \lambda_i, \delta_{i\theta} \right) - \cov \left( \gamma \ln Q_{\theta} , \delta_{i\theta} \right) + \cov\left( \varepsilon_{i\theta}, \delta_{i\theta} \right) 
=0
\end{align}
which holds true because of Assumption \ref{ortho} and the nonrandomness of $\lambda_{i}$.

To identify the elasticity parameter $\gamma$, we return to the equilibrium equation (\ref{eeq}), which is reiterated as follows:
\begin{align}
\ln S_{it} = \phi \ln \lambda_i + \kappa \left({\tau} - \ln Q_t \right) + \kappa \ln Z_{it} + \phi (\varepsilon_{it} + \gamma \delta_{it} )
\label{regppmod}
\end{align}
where $\phi = \frac{1}{1-\gamma {\omega}}$ and $\kappa = \frac{\gamma}{1-\gamma {\omega}}$.\footnote{
Note that $\phi=\frac{1}{1-\gamma \omega}=1+\kappa\omega$, which is estimable via $\hat{\kappa}$ and $\hat{\omega}$, turns out to be equal to the exchange rate pass-through (ERPT), i.e., the elasticity of local currency import prices with respect to the local currency price of foreign currency.
More details are appended to the \ref{appdx4}.
}
As mentioned previously, there is also no endogeneity problem for this regression equation.
By way of fixed effects or double-demeaned regression, $\kappa$ can be identified.
The double-demeaned version of Equation (\ref{regppmod}) is described as follows:
\begin{align}
\left[ \ln S_{it} \right]= \kappa \left[ \ln Z_{it} \right] +  \phi \left[ \varepsilon_{it} + \gamma \delta_{it} \right]
\label{regppdd}
\end{align}
Finally, trade elasticity is identified by utilizing the estimates of regressions (\ref{regsupply}) and (\ref{regppdd}) in the following manner:
\begin{align}
\hat{\sigma} = 1-\hat{\gamma} = 1 - \frac{\hat{\kappa}}{1+\hat{\kappa} \hat{\omega} }
\label{fourth}
\end{align}
Since the error terms of (\ref{regsupply}) and (\ref{regppdd}) seem to be correlated, we use SUR (seemingly unrelated regressions) to efficiently estimate the elasticity via (\ref{fourth}).

\section{Empirical analysis}
In addition to the identification strategy presented above, to estimate the trade elasticity from the demand shares and exchange rates, we examine two further alternatives, which we describe in detail in \ref{appdx1}.
The first is to directly apply Feenstra's method, which we refer to as FM, where the trade elasticity is identified by (\ref{solfeenstra}).
The second, which we refer to as IIV, applies the implicit instrumental variable of Feenstra's method (as defined thereof) to the explanatory variable of (\ref{regwithz}).
Together with these two alternatives, we examine our strategy, referred to as SUR, which performs regression analyses for the supply equation (\ref{regsupply}) and the equilibrium equation (\ref{regppdd}) simultaneously to obtain the trade elasticity by (\ref{fourth}).
We examine these three alternatives (FM, IIV, and SUR) to estimate the trade elasticities for commodities whose physical quantities are available.
Note that $\gamma$ can be estimated via Equation (\ref{regmain}) via fixed effects while the exchange rate $Z_{it}$ is applied as an instrument to fix the endogenous explanatory variable $\ln P_{it}$, where $P_{it}=Z_{it}V_{it}/X_{it}$, if quantities $X_{it}$ are available.
We use this estimator, which we refer to as IVFE, as the benchmark for assessing the abovementioned three alternatives.
In addition, we note that the import values $V_{it}$ used in the benchmark estimations do not include tariffs since the analyses are intended for the comparative assessment of alternative means applicable to services.

%%%%%%%%%%%%%%%%%%%%%%%%%%%%%%%%
\newcolumntype{i}{D{. }{}{0}}
\begin{table}[t]
\begin{threeparttable}
\caption{Benchmark IVFE estimation in the presence of quantity.} \label{tab_bench}
\newcolumntype{.}{D{.}{.}{9}}
\newcolumntype{*}{D{.}{.}{2}}
\newcolumntype{!}{D{.}{.}{8}}
\begin{tabularx}{\linewidth}{r . . * ! ! }
\hline\noalign{\smallskip}
& \multicolumn{1}{c}{FE} & \multicolumn{4}{c}{FE(IV)} \\
\cmidrule(r){2-2}\cmidrule(r){3-6}
\multicolumn{1}{r}{commodity} 
&\multicolumn{1}{c}{$\hat{\sigma}$ (s.e.)}
&\multicolumn{1}{c}{$\hat{\sigma}$ (s.e.)}
&\multicolumn{1}{c}{1F\tnote{*1}}
&\multicolumn{1}{c}{overid.\tnote{*2}}
&\multicolumn{1}{c}{endog.\tnote{*3}}
\\ \hline\noalign{\smallskip}
\input{Tab_IVFE2.tex} \\\hline
\end{tabularx}
\begin{tablenotes}
\footnotesize
\item[*1] First-stage (Cragg-Donald Wald) F statistic for 2SLS FE estimation.
The rule of thumb to reject the hypothesis that the explanatory variable is correlated with the instrument is for this to exceed 10.
\item[*2] Overidentification test by Sargan statistic.
Rejection of the null indicates that the instruments are correlated with the residuals.
Blank indicates that a secondary instrument could not be found.
\item[*3] Endogeneity test by Davidson-MacKinnon F statistic.
Rejection of the null indicates that the instrumental variable fixed effects estimator should be employed.
\end{tablenotes}
\end{threeparttable}
\end{table}
%%%%%%%%%%%%%%%%%%%%%%%%%%%%%%%

\subsection{Data}
Our main strategy for identifying the supply parameter via Equation (\ref{regsupply}) relies on the cross-sectional variety of exporters of a commodity.
For this reason, we examine mainly the imports of food, whose origins are generally diverse.
Specifically, we gather Japan's monthly import value and quantity data from \cite{boeki} spanning from 2008 January to 2022 December for 11 commodities, namely, coffee, beer, wine, beef, chicken, salmon, blue fish, flat fish, crabs, shrimp, and rocks.
More details are appended to \ref{appdx2}.
The monthly exchange rates pertaining to all partner countries involved are drawn from \citet{fx}.

\subsection{Results}
First, we estimate benchmark trade elasticities for the 11 commodities mentioned above via IVFE.
The results are shown in Table \ref{tab_bench}.
The instrumental variables we apply to the endogenous variable $\ln P_{it}$ of Equation (\ref{regmain}) are $\ln Z_{it}$ and $Z_{it}$ whenever they pass the overidentification test.
Ultimately, where the overidentification test statistic is blank, only the primary instrument $\ln Z_{it}$ was applied.
According to the endogeneity test statistics, we employ fixed effects (FE) estimators for blue fish and rocks, whereas instrumental variable estimators are employed for the remaining commodities.
We excluded beer, crabs, shrimp and flat fish from further analyses since negative elasticity does not support the microfoundation of the model we support.
The employed benchmark elasticities are presented in Table \ref{tab_sum} (IVFE column).
%%%%%%%%%%%%%%%%%%%%%%%%%%%%%%%%%%%%%%
\begin{table}[t]
\begin{threeparttable}
\caption{Estimation via supply and equilibrium equations.} \label{tab_sur}
\newcolumntype{.}{D{.}{.}{10}}
\begin{tabularx}{\linewidth}{l . . . .}
\hline\noalign{\smallskip}
& \multicolumn{1}{c}{SUR} & \multicolumn{2}{c}{Equation (\ref{regsupply})\tnote{*2}}& \multicolumn{1}{c}{Equation (\ref{regppdd})} \\
\cmidrule(r){2-2}\cmidrule(r){3-4}\cmidrule(r){5-5} %\noalign{\smallskip}
\multicolumn{1}{c}{Sector} 
&\multicolumn{1}{c}{$\hat{\sigma}$~(s.e.)\tnote{*1} } 
&\multicolumn{1}{c}{$\hat{\omega}$~(s.e.)} 
&\multicolumn{1}{c}{const.~(s.e.)} 
&\multicolumn{1}{c}{$\hat{\kappa}$~(s.e.)}
\\ \hline \noalign{\smallskip}
\input{Tab_SUR2.tex} \\\hline
\end{tabularx}
\begin{tablenotes}
\footnotesize
\item[*1] Delta method standard errors.
\item[*2] The point of normalization was selected at the final year, 2020. 
\end{tablenotes}
\end{threeparttable}
\end{table}
%%%%%%%%%%%%%%%%%%%%%%%%%%%%%%%%%%%%%%%%

Table \ref{tab_sur} summarizes the results of our %identification
strategy to identify the demand parameter (i.e., trade elasticity) from the supply and equilibrium equations.
Regarding the supply equation (\ref{regsupply}), the point of normalization $t=\theta$, where we perform the cross-sectional regression, is arbitrary.
Here, we select the point of normalization as the minimum RSS of the cross-sectional regression.
The second column of Table \ref{tab_sur} shows the results of performing independent regressions for the supply equation (\ref{regsupply}) at the corresponding point of normalization.
The third column shows the results of performing independent regressions for the equilibrium equation (\ref{regppdd}).
Finally, the first column (entitled SUR) shows the results of performing seemingly unrelated regression for the system of equations (\ref{regsupply}) and (\ref{regppdd}).
The estimated trade elasticities are reiterated in Table \ref{tab_sum} (SUR).

\begin{figure}[t]
\centering
\includegraphics[width=0.49\textwidth]{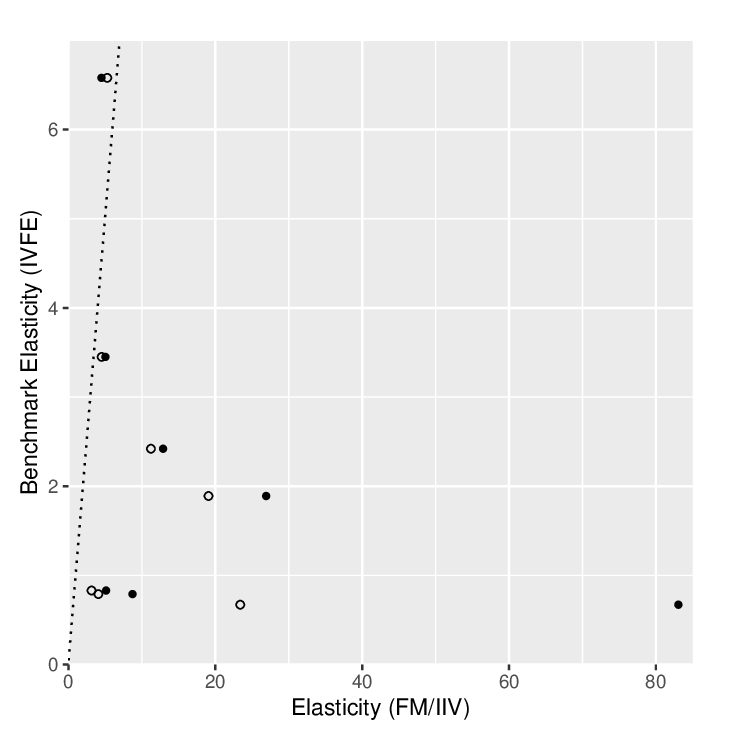}
\includegraphics[width=0.49\textwidth]{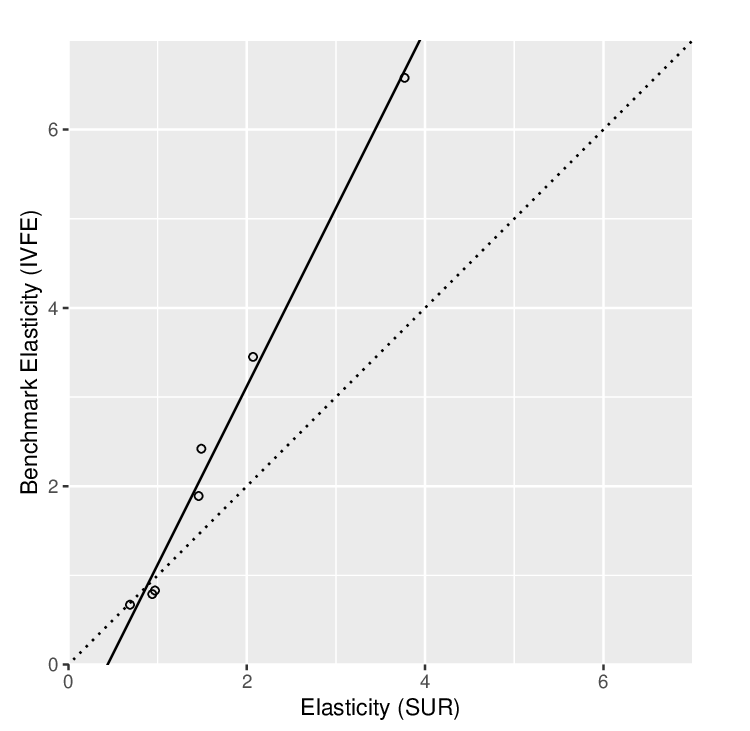}
\caption{
For both figures, the vertical axis corresponds to the benchmark elasticity (estimated via IVFE) of commodities with relevant (nonnegative) point estimates, namely, salmon, coffee, beef, blue fish, rocks, chicken, and wine, in descending order.
The horizontal axis corresponds to the elasticity estimate by means of the alternatives, namely, FM (closed dots, left), IIV (open dots, left) and SUR (right).
The dotted line is the 45-degree line.
} \label{fig_assess}
\end{figure}

Finally, Table \ref{tab_sum} and Figure \ref{fig_assess} summarize the point estimates by the three alternatives (FM, IIV, SUR) with the benchmark estimates (IVFE) of the trade elasticities.
The estimates obtained via FM and IIV are transcribed from Table \ref{tab_fmiiv}, which appears in \ref{appdx1}.
FM and IIV are very close in finding the benchmark elasticity for two commodities, salmon and coffee, whereas SUR was close for all seven commodities.
The solid line in Figure \ref{fig_assess} (right) is the regression line specified below:
\begin{align}
\text{Benchmark Elasticity (IVFE)} = - \underset{(0.176)}{0.879} + \underset{(0.093)}{2.000}\times \text{Elasticity (SUR)} 
\label{correction}
\end{align}
Standard errors are in parentheses, and the adjusted $R^2 = 0.987$.
Note that the trade elasticity for services can be estimated via SUR and further corrected for increased precision via this formula.

%%%%%%%%%%%%%%%%%%%%%%%%%%%%%%%%%%%%%%
%%%%%%%%%%%%%%%%%%%%%%%%%%%%%%%%%%%
\begin{table}[t]
\begin{threeparttable}
\caption{Summary of point estimates by various methods.} \label{tab_sum} 
\newcolumntype{.}{D{.}{.}{10}}
%\center
\begin{tabularx}{\linewidth}{r . . . .}
\hline\noalign{\smallskip}	
& \multicolumn{1}{c}{IVFE} & \multicolumn{1}{c}{FM}& \multicolumn{1}{c}{IIV}& \multicolumn{1}{c}{SUR}  \\
\cmidrule(r){2-2}\cmidrule(r){3-3}\cmidrule(r){4-4}\cmidrule(r){5-5}
\multicolumn{1}{c}{commodity}&\multicolumn{1}{c}{$\hat{\sigma}$ (s.e.)}&\multicolumn{1}{c}{$\hat{\sigma}$ (s.e.)}&\multicolumn{1}{c}{$\hat{\sigma}$ (s.e.)}&\multicolumn{1}{c}{$\hat{\sigma}$ (s.e.)}\\ \hline \noalign{\smallskip}
\input{Tab_Result2.tex} \\\hline
\end{tabularx}
\begin{tablenotes}
\footnotesize
\item[*1] Estimation result for FM and IIV are blanked due to anomalies.
\end{tablenotes}
\end{threeparttable}
\end{table}
%%%%%%%%%%%%%%%%%%%%%%%%%%%%%%%%%%%%%

\subsection{Application to services}
We employ services imports data from \citet{wtostats}.
Japan was selected as the reporting economy, and yearly import values (in million USD) were drawn for 1) commercial services, 2) construction, 3) financial services, 4) goods-related services, 5) insurance and pension services, 6) telecommunications, computer, and information services, 7) transport, and 8) travel, spanning the timeline from 2005--2020.
The corresponding yearly average exchange rates (in LCC/JPY) are obtained from \citet{fx}.
The normalization point was selected at the last period ($\theta = 2020$) since the number of exporting countries (hence, the size of the sample) increased over time.

Table \ref{tab_service} shows the results of estimating the trade
elasticities of the abovementioned eight different services, by means
of SUR, on the basis of regression equations (\ref{regsupply}) and (\ref{regppdd}).
Since the point estimates are generally around unity, we opted not to use the correction formula (\ref{correction}).
Notably, in this table, $\hat{\sigma} \approx 1 - \hat{\kappa}$, which indicates the ineffectiveness of $\hat{\omega}$ or the supply equation (\ref{regsimul2}), which can potentially cause endogeneity.
Overall, the elasticity estimates turned out to be near unity, meaning that the Arminton aggregators for services are generally Cobb-Douglas.
Under a Cobb-Douglas aggregator, the import demand value (in terms of the local currency) is unresponsive to changes in price.
Furthermore, we note that 4) goods-related services and 5) insurance and pension services, among other services, are internationally complementary (i.e., $\hat{\sigma} <1$), meaning that the import demand quantity is relatively unresponsive to changes in price.
%%%%%%%%%%%%%%%%%%%%%%%%%%%%%%%%%%%%%%%
\begin{table}[t]
\begin{threeparttable}
\caption{Trade elasticity estimates for services.} \label{tab_service}
\newcolumntype{.}{D{.}{.}{10}}
\begin{tabularx}{\linewidth}{l . . . .}
\hline\noalign{\smallskip}	
& \multicolumn{1}{c}{SUR} & \multicolumn{2}{c}{Equation (\ref{regsupply})\tnote{*2}}& \multicolumn{1}{c}{Equation (\ref{regppdd})} \\
\cmidrule(r){2-2}\cmidrule(r){3-4}\cmidrule(r){5-5} 
\multicolumn{1}{c}{commodity} 
&\multicolumn{1}{c}{$\hat{\sigma}$~(s.e.)\tnote{*1} } 
&\multicolumn{1}{c}{$\hat{\omega}$~(s.e.)} 
&\multicolumn{1}{c}{const.~(s.e.)} 
&\multicolumn{1}{c}{$\hat{\kappa}$~(s.e.)}
\\ \hline \noalign{\smallskip}
\input{Tab_Service2.tex} \\\hline
\end{tabularx}
\begin{tablenotes}
\footnotesize
\item[*1] Delta method standard errors.
\item[*2] The point of normalization was selected at the final year, 2020. 
\end{tablenotes}
\end{threeparttable}
\end{table}
%%%%%%%%%%%%%%%%%%%%%%%%%%%%%%%%%%%%%%

\section{Concluding Remarks}
\citet{oecd_benz} presents a summary of the estimates of the trade elasticities for services provided by the aforementioned sources \citep[i.e., ][]{oecd, christen, egger, blank}, which base their analyses on the framework of monopolistic competition and firm data.
\citet{oecd_benz} also provides the simple average of these estimates, which we transcribe as follows:
3.67 (communication), 3.21 (business services), 2.54 (financial services), 2.77 (insurance), 3.39 (transport).
These figures indicate that the intranational firm-wise variety of services is highly elastic, i.e., the services can easily be substituted for each other.
On the other hand, our estimates for service trade elasticities summarized in Table \ref{tab_service} indicate substantial complementarity in the international variety of services.
Our findings seem to accord with our understanding of the elasticity of substitution.

\subsubsection*{Acknowledgements}
The authors would like to thank the anonymous reviewers and the associate editor of the journal for helpful comments on the earlier version of the paper.\\
%JSPS Kakenhi Grant numbers: 19H04380, 20K22139 %\\
%The authors declare that they have no conflicts of interest.

\def\thesection{Appendix 1}
\section{\label{appdx1}}
\subsubsection*{
Feenstra's method and extensions
}
Here, we consider estimating $\gamma$ via Equation (\ref{regdd2}) via Feenstra's method to address the endogeneity problem without finding external instruments.
Below, we write the main (endogenous) regression (\ref{regdd2}) with
obvious notational changes and the corresponding reverse-cause
equation.\footnote{
Feenstra's method is designed to provide a consistent estimate of substitution elasticity via a demand equation such as that of (\ref{regdd}); thus, the reverse-cause equation usually corresponds to a supply equation.
Moreover, instead of double-demeaning the variables, Feenstra's method
takes time differences and item differences from a reference item to eliminate time- and item-specific effects.
}
\begin{align}
s_{it} =\gamma z_{it} + \mu_{it},
&&
z_{it} = \rho s_{it} + \nu_{it}
\end{align}
Following Feenstra's method, we multiply the zero-mean error terms of the above equations to obtain the following equation:
\begin{align}
U_{it} = \alpha_1 W_{1it} + \alpha_2 W_{2it} + \xi_{it}
\label{feenstra}
\end{align}
where $\xi_{it}=-\mu_{it}\nu_{it}/\gamma$, $W_{1it}=(s_{it})^2$, $W_{2it}=s_{it}z_{it}$, $\alpha_1 = -\rho/\gamma$, $\alpha_2 = (1+\gamma \rho)/\gamma$ and $U_{it}=z_{it}z_{it}$.
Then, we take the temporal average of (\ref{feenstra}) as follows:
\begin{align}
\left< U_i \right> = \alpha_1 \left< W_{1i} \right> + \alpha_2 \left< W_{2i} \right> + \left<\xi_{i} \right>
\label{feenstra2}
\end{align}
where $\left<U_{i}\right>=\sum_{t=1}^{T} U_{it}/T$ and so forth for the full-panel case.
Assume that $\plim \left< \xi_i \right> =0$, in which case we can
assert that $\plim \left< W_{ni}\right> \left<\xi_i \right> = (\plim
\left< W_{ni} \right>)(\plim \left< \xi_i \right>)=0$ for $n=1,2$, or
the consistency of the least squares estimators of regression
(\ref{feenstra2}), where we can solve for $\hat{\gamma}$ and ${\hat\rho}$ via the following formula:
\begin{align}
\hat{\gamma} = \frac{-\hat{\alpha}_2 \pm \sqrt{(\hat{\alpha}_2 )^2 + 4 \hat{\alpha}_1}}{2 \hat{\alpha}_1},
&&
\hat{\rho} = \frac{\hat{\alpha}_2 \mp \sqrt{( \hat{\alpha}_2 )^2 + 4 \hat{\alpha}_1}}{2}
\label{solfeenstra}
\end{align}
In practice, with unbalanced panel data, Equation (\ref{feenstra2}) is estimated via WLS (weighted least squares) regression, using $T_i/v_i$ as weights where
$T_i$ and $v_i$ denote the temporal size and the variance of the
residuals of the $i$th panel, respectively.

Note that our key assumption (i.e., $\plim \left< \xi_i \right> =0$) is equivalent to the following:
\begin{align}
\plim_{T \to \infty} \frac{1}{T}\sum_{t=1}^T \mu_{it}\nu_{it}  =0 
\end{align}
That is, in Feenstra's method, we assume asymptotic uncorrelation (i.e., independence) between $\mu_i$ and $\nu_i$ for all $i$.
This is the point from which our second strategy stems.
By this assumption, we can use $\hat{\rho}$ from the first strategy to recover the residual $\hat{\nu}_{it}$, which must be uncorrelated with the error term of the main demand equation $\mu_{it}$ and yet is correlated with the response variable $z_{it}$ by construction.
The residual $\hat{\nu}_{it}$ must therefore be a relevant instrument for estimating the main demand equation (\ref{regdd2}) or (\ref{regwithz}).
We call this instrument ($\hat{\nu}$) the IIV (implicit instrumental variable) of Feenstra's method.

Table \ref{tab_fmiiv} summarizes the results obtained via Feenstra's method (FM) and an extension that uses the corresponding implicit instrumental variables (IIVs).
Specifically, our primary IIV is obtained via the following protocol, which uses $\hat{\rho}$ from Equation (\ref{solfeenstra}):
\begin{align}
\hat{\nu} = [\ln Z_{it}] - \hat{\rho} [\ln S_{it}] 
\end{align}
We apply the instrumental variable $\hat{\nu}$ and its derivatives
onto the explanatory variable $[\ln Z_{it}]$ of Equation (\ref{regdd2}) whenever it passes the overidentification test.
{Note that by derivatives of $\nu_{it}$, we mean $\ln \hat{\nu}_{it}$, its forward and backward lags, and its first differences.}
Ultimately, where the overidentification test statistic is blank, only the primary instrument $\hat{\nu}_{it}$ was applied.
As anticipated, both alternative methods provide similar results, except for flat fish.
The estimated trade elasticities are presented in Table \ref{tab_sum}
(columns FM and IIV).
\begin{table}[t]
\begin{threeparttable}
\caption{Applying Feenstra's method.} \label{tab_fmiiv}
\newcolumntype{.}{D{.}{.}{9}}
\newcolumntype{,}{D{.}{.}{3}}
\newcolumntype{*}{D{.}{}{0}}
\begin{tabularx}{\linewidth}{r . . * . , }
\hline\noalign{\smallskip}	
& \multicolumn{1}{c}{FM} & \multicolumn{4}{c}{IIV} \\
\cmidrule(r){2-2}\cmidrule(r){3-6} 
\multicolumn{1}{r}{commodity} &\multicolumn{1}{c}{$\hat{\sigma}$ (s.e.)} &\multicolumn{1}{c}{$\hat{\sigma}$ (s.e.)} &\multicolumn{1}{c}{1F\tnote{*1}} &\multicolumn{1}{c}{overid.\tnote{*2}} &\multicolumn{1}{c}{endog.\tnote{*3}}  \\ \hline \noalign{\smallskip}
\input{Tab_IIV2.tex} \\\hline
\end{tabularx}
\begin{tablenotes}
\footnotesize
\item[*1] First-stage (Cragg-Donald Wald) F statistic for 2SLS FE estimation.
The rule of thumb to reject the hypothesis that the explanatory variable is correlated with the instrument is for this to exceed 10.
\item[*2] Overidentification test by Sargan statistic.
Rejection of the null indicates that the instruments are correlated with the residuals.
Blank indicates that a secondary instrument could not be found.
\item[*3] Endogeneity test by Davidson-MacKinnon F statistic.
Rejection of the null indicates that the instrumental variable fixed effects estimator should be employed.
\item[*4] Estimation resulted in anomalies.
\end{tablenotes}
\end{threeparttable}
\end{table}

\def\thesection{Appendix 2}\section{\label{appdx2}}
Table \ref{tab_hs} shows the HS codes eligible for each tangible commodity.
The size of the sample is $T=180$ in all cases for the 11 tangible commodities.
The size of panel $N$ is restricted to the number of countries that have more than 160 observations out of the full 180, except for beef.
To ensure the strength of our primary instrument, we use data spanning from 2006--2020 for chicken and from 2001--2015 for beef and restrict the panel to those with more than 100 observations.
For the same purpose, we excluded South American wines from the wine data.
\begin{center}
\begin{ThreePartTable}
\begin{TableNotes}
\vspace{-1em}
\footnotesize
\begin{itemize}
\item[Beef ] Carcasses, cuts, loin, kata, ude, momo, and belly cuts that are fresh, chilled or frozen.
\item[Beer ] Beer made from malt.
\item[Blue fish ] Mackerel, Jack \& Horse mackerel, sardines, anchovies, pacific saury that are fresh, chilled or frozen.
\item[Chicken ] Meat, legs, cuts, and offal that are fresh, chilled or frozen.
\item[Coffee ] Coffee, not roasted, not decaffeinated.
\item[Crabs ] King, snow, swimming, horsehair, and mitten crabs that are fresh, chilled or frozen.
\item[Flat fish ] Hirame, halibut, sole, and turbots, and fillets thereof, which are fresh, chilled or frozen.
\item[Salmon ] Red, silver, pacific, Atlantic \& Danube salmon, and Salmonidae, which are fresh, chilled or frozen.
\item[Shrimps ] Shrimps, prawns, ebi, and crustaceans that are fresh, chilled, frozen, or smoked.
\item[Rocks ] Setts, curbstones and flagstones of natural stone.
\item[Wine ] Wine in various containers and grape must.
\end{itemize}
\end{TableNotes}
\begin{longtable}{p{0.08\textwidth} p{0.85\textwidth}}
\caption{Commodities and HS codes. }\label{tab_hs} \\
\hline \noalign{\smallskip}
 & HS Code \\
\hline
\endfirsthead
\multicolumn{2}{l}{Continued.} \\
\hline \noalign{\smallskip}
 & HS Code
\endhead
\endfoot
\insertTableNotes
\endlastfoot
%------------------------
\input{Tab_HScodes.tex}
\end{longtable}
\end{ThreePartTable}
\end{center}

\def\thesection{Appendix 3}
\section{\label{appdx3}}
\subsubsection*{
STRI elasticity of imports
}

The OECD Service Trade Restrictiveness Index (STRI) provides comparable information on regulations affecting trade in services across 50 countries and 22 sectors from 2014 \citep{stri}.
Our model presented in Section \ref{sec_model} uses exchange rates as the only observable variable to estimate trade elasticities, whereas other exogenous disturbances are incorporated as part of demand shocks.
In this section, we examine how the STRI data could contribute to the scraping of relevant disturbances and increase the efficiency of the elasticity estimation.
To illustrate the modifications, let us first decompose the disturbance term $\varepsilon_{it}$ of the demand equation (\ref{regsimul1}) into two parts.
Below are the modified demand and supply equations:
\begin{align}
\ln S_{it} &= \ln \lambda_i  -\gamma \ln Q_t + \gamma \left( \ln Z_{it}  + \ln \pi_{it}  \right) + \epsilon_{it} +  \eta \ln R_{it}  \label{dem2} 
\\
\ln \pi_{it} &= \tau + \omega \ln S_{it}  +  \delta_{it}  \tag{\ref{regsimul2}}
\end{align}
Here, $\varepsilon_{it} =  \epsilon_{it} + \eta \ln R_{it}$ where $R_{it}$ denotes the restrictiveness index (STRI) and where $\eta$ denotes the STRI elasticity of service imports (value).\footnote{
This is possible on the ground that $[\ln S_{it}]=[\ln V_{it} - \ln \sum_{i=1}^N V_{it} ]={[\ln V_{it}]}$.
}
According to \citet{stri}, the STRIs are composite indices with values between zero and one, with zero representing an open market and one representing a market completely closed to foreign service providers.
The weighting scheme for the STRI is derived from an online survey that collects binary (yes/no) answers, while measures that have numerical answers are broken down on thresholds to which binary scores are applied.
We may therefore be convinced that STRIs have nothing to do with exchange rates or with remaining external demand shocks $ \epsilon_{it}$.
In summary, we make the following assumption:
\begin{assumption}\label{ortho2}
The supply shocks $\delta_{it}$, demand shocks $\epsilon_{it}$, and (growth rate of) trade restrictiveness $\ln R_{it}$ are mutually independent, i.e.,
\begin{align*} 
\cov (\epsilon_{it}, \delta_{it}) = \cov (\epsilon_{it}, \ln R_{it}) = \cov (\delta_{it}, \ln R_{it}) =0
\end{align*}
\end{assumption}
Combining the demand and supply equations (\ref{dem2}) and (\ref{regsimul2}), we have the following (modified) equilibrium equation:
\begin{align*}
\ln S_{it} 
= \phi \ln \lambda_i + \kappa {\tau} - \kappa \ln Q_t + \kappa \ln Z_{it} + \mu \ln R_{it} + (\phi  \epsilon_{it} + \phi \gamma \delta_{it}) 
\end{align*}
where $\mu = \frac{\eta}{1-\gamma \omega}$, $\kappa = \frac{\gamma}{1-\gamma \omega}$, and $\phi = \frac{1}{1-\gamma \omega}$.
By virtue of Assumptions \ref{ortho} and \ref{ortho2}, the following double-demeaned equilibrium equation must be free of endogeneity:\footnote{Moreover, the multivariate regression (\ref{ddeqeq}) must be free of multicollinearity.}
\begin{align}
[\ln S_{it}]
= \kappa [\ln Z_{it}] + \mu [\ln R_{it}] + [\phi \epsilon_{it} + \phi \gamma \delta_{it} ]
\label{ddeqeq}
\end{align}
Normalizing the price at $t=\theta$, i.e., $(\ln Z_{i\theta} + \ln \pi_{i\theta})=0$, renders the following:
\begin{align}
\ln S_{i\theta} &= \ln \lambda_{i} -\gamma \ln Q_{\theta} + \epsilon_{i\theta} + \eta \ln R_{i\theta} \notag
\\
-\ln Z_{i\theta} &= \tau + \omega \ln S_{i\theta} + \delta_{i\theta} \label{normal2}
\end{align}
Then, we know that regression (\ref{normal2}) is endogeneity-free, in light of the following expansion:
\begin{align*}
\cov(\ln S_{i\theta}, \delta_{i\theta}) 
= \cov( \ln \lambda_{i} -\gamma \ln Q_{\theta} + \epsilon_{i\theta} + \eta \ln R_{i\theta},  \delta_{i\theta})
=0
\end{align*}
which holds true by Assumption \ref{ortho2} and the nonrandomness of $\lambda_i$.
The trade elasticity $\sigma$ and the STRI elasticity of service imports $\eta$ are therefore identified by the estimates of regressions (\ref{ddeqeq}) and (\ref{normal2}) in the following manner:
\begin{align*}
\hat{\sigma} =1 - \frac{\hat{\kappa}}{1+\hat{\kappa} \hat{\omega}},
&&
\hat{\eta} = \frac{\hat{\mu}}{1+ \hat{\kappa} \hat{\omega}}
\end{align*}
Since the error terms of regressions (\ref{ddeqeq}) and (\ref{normal2}) seem to be correlated, we use SUR to efficiently estimate the elasticities $\sigma$ and $\eta$.

We draw the STRI data from \citet{stri}.
Selecting Japan as the partner country, yearly values of the STRI were drawn for all service sectors available.
We find equivalent sectors of STRI data in the WTO's services imports data: 2) construction (construction), 3) financial services (commercial banking), 5) insurance and pension services (insurance), 6) telecommunications, computer, and information services (telecom and computer), and 7) transport (air transport, maritime transport, road freight transport, and rail freight transport).
However, we could not find relevant sectors for 1) commercial services, 4) goods-related services, and 8) travel.
We therefore performed SUR to estimate the trade elasticity and the STRI elasticity of imports at the same time for services of ID numbers 2), 3), 5), 6), and 7).
The results are summarized in Table \ref{tab_service2}.

The estimates for trade elasticity are different from those reported in Table \ref{tab_service}, mainly because of the different time ranges of observations (2014--2020) and the range of countries covered in the STRI database.
For the estimates for the STRI elasticity of service imports, we find statistical significance (with the correct sign) for 3) financial services, where the STRI of the commercial banking sector is applied.
While statistically insignificant, the STRI elasticity of imports for 2) Construction has an opposite sign rather than what we can anticipate from our intuition that trade restrictions should have negative effects on import values.
The peculiarity of the construction sector can also be recognized in its mode structure studied in \citet{eurostat}, with a 90\% share of Mode 3 (commercial presence) and a 10\% share of Mode 4 (movement of natural persons). 
%\footnote{ %have 90\% and 10\% shares, respectively.\footnote{
%Mode 3 (commercial presence): mode of supply by a service supplier of one country through a commercial presence in the territory of another country.
%Mode 4 (movement of natural persons): mode of supply by a service supplier of one country through the presence of natural persons of that country in the territory of any other country.}
%%%%%%%%%%%%%%%%%%%%%%%%%%%%%%%%%%%%%
\begin{table}[t]
\begin{threeparttable}
\caption{Estimation of trade elasticity and STRI elasticity of import.} \label{tab_service2}
\newcolumntype{.}{D{.}{.}{8}}
\begin{tabularx}{\linewidth}{r..  ...}
\hline\noalign{\smallskip}	
 & \multicolumn{2}{c}{SUR} & \multicolumn{1}{c}{Equation (\ref{normal2})\tnote{*2}} & \multicolumn{2}{c}{Equation (\ref{ddeqeq})} \\
\cmidrule(l){2-3}\cmidrule(l){4-4}\cmidrule(l){5-6}
ID & \multicolumn{1}{c}{$\hat{\sigma}$~~(s.e.)\tnote{*1}} & \multicolumn{1}{c}{$\hat{\eta}$~~(s.e.)\tnote{*1}}
& \multicolumn{1}{c}{$\hat{\omega}$~~(s.e.)}& \multicolumn{1}{c}{$\hat{\kappa}$~~(s.e.)}& \multicolumn{1}{c}{$\hat{\mu}$~~(s.e.)}
\\ \hline \noalign{\smallskip}
\input{Tab_STRI.tex} \\\hline
\end{tabularx}
\begin{tablenotes}
\footnotesize
\item[*1] Delta method standard errors.
\item[*2] The estimates for constant are suppresed.
The point of normalization was selected at the final year, 2020. 
\end{tablenotes}
\end{threeparttable}
\end{table}
%%%%%%%%%%%%%%%%%%%%%%%%%%%%%%%%%%%%%

\def\thesection{Appendix 4}
\section{\label{appdx4}}
%\subsubsection*{ERPT}
According to \citet{GK, CG}, exchange rate pass-through (ERPT) is defined as the elasticity of local currency import prices $P = Z \pi$ with respect to the exchange rates $Z$ (i.e., local currency price of foreign currency), where $\pi$ denotes the foreign currency import prices, as defined previously.
ERPT can therefore be expanded as follows:
\begin{align*}
\text{ERPT} = \frac{\partial \ln P}{\partial \ln Z} &= \frac{\partial \ln Z \pi}{\partial \ln Z}
= 1 + \frac{\partial \ln \pi}{\partial \ln Z} \\
&= \frac{\partial \ln P}{\partial \ln S}\frac{\partial \ln S}{\partial \ln \pi}\frac{\partial \ln \pi}{\partial \ln Z}=\frac{1}{\gamma \omega}\frac{\partial \ln \pi}{\partial \ln Z}
\end{align*}
where we used $\frac{\partial S}{\partial P} = \gamma$ and $\frac{\partial \pi}{\partial S} = \omega$ from  (\ref{regmain}) and (\ref{regsimul2}) for the last identity.
Thus, by eliminating $\frac{\partial \ln \pi}{\partial \ln Z}$ from the above equations, we arrive at $\text{EPRT}=\frac{1}{1-\gamma\omega}=\phi$.

\bibliographystyle{spbasic_x}
{\raggedright
\bibliography{bibfile}
}

\end{document}

%% file: Tab_IVFE2.tex
Coffee	&	1.614	~(0.071)	&	3.446	~(0.500)	&	36.84	&	0.52	~(0.469)	&	16.69	~(0.000)	\\
Beef	&	3.071	~(0.122)	&	2.424	~(0.351)	&	50.98	&	0.30	~(0.582)	&	4.06	~(0.044)	\\
Blue fish	&	1.892	~(0.106)	&	1.594	~(0.238)	&	83.00	&	0.17	~(0.680)	&	1.88	~(0.170)	\\
Chicken	&	1.649	~(0.068)	&	0.792	~(0.363)	&	15.69	&	0.48	~(0.489)	&	6.85	~(0.009)	\\
Rocks	&	0.829	~(0.101)	&	0.795	~(0.893)	&	14.26	&			&	0.00	~(0.988)	\\
Salmon	&	1.809	~(0.113)	&	6.575	~(1.287)	&	11.87	&	0.79	~(0.373)	&	29.42	~(0.000)	\\
Wine	&	1.165	~(0.027)	&	0.671	~(0.184)	&	27.90	&	1.23	~(0.267)	&	8.39	~(0.004)	\\
Cotton	&	1.179	~(0.171)	&	1.035	~(0.725)	&	11.24	&	3.02	~(0.082)	&	0.04	~(0.839)	\\
Beer	&	1.270	~(0.051)	&	-0.701	~(0.796)	&	16.79	&			&	9.43	~(0.002)	\\
Crabs	&	0.799	~(0.075)	&	-1.313	~(0.590)	&	12.94	&	0.02	~(0.899)	&	23.03	~(0.000)	\\
Shrimps	&	1.538	~(0.050)	&	-2.551	~(1.168)	&	8.94	&	0.90	~(0.343)	&	35.56	~(0.000)	\\
Flat fish	&	1.411	~(0.129)	&	-1.796	~(1.116)	&	14.68	&	1.79	~(0.181)	&	11.36	~(0.001)	

%% file: Tab_SUR2.tex
Coffee	&	2.070	~(0.441)	&	0.554	~(0.271)	&	2.238	~(1.402)	&	-0.672	~(0.126)	\\
Beef	&	1.494	~(0.183)	&	-0.654	~(0.093)	&	-5.444	~(0.524)	&	-0.729	~(0.242)	\\
Blue fish	&	1.461	~(0.171)	&	-0.846	~(0.504)	&	-5.165	~(1.883)	&	-0.757	~(0.403)	\\
Chicken	&	0.938	~(0.088)	&	-0.596	~(0.221)	&	-4.082	~(1.206)	&	0.060	~(0.110)	\\
Rocks	&	0.965	~(0.125)	&	-1.756	~(0.691)	&	-6.178	~(2.631)	&	0.033	~(0.143)	\\
Salmon	&	3.772	~(2.572)	&	0.720	~(0.250)	&	0.283	~(1.202)	&	-0.925	~(0.144)	\\
Wine	&	0.690	~(0.154)	&	0.384	~(0.236)	&	-1.491	~(1.369)	&	0.352	~(0.176)	\\
Cotton	&	-6.721	~(13.69)	&	-1.128	~(0.025)	&	-5.502	~(0.063)	&	0.795	~(0.146)	\\
Beer	&	0.195	~(0.606)	&	-0.880	~(0.436)	&	-5.347	~(1.771)	&	0.471	~(0.179)	\\
Crabs	&	1.132	~(0.021)	&	-1.224	~(0.262)	&	-5.167	~(1.223)	&	-0.158	~(0.030)	\\
Shrimps	&	1.110	~(0.020)	&	0.898	~(0.250)	&	3.191	~(1.265)	&	-0.100	~(0.020)	\\
Flat fish	&	0.287	~(0.493)	&	-0.989	~(0.438)	&	-5.807	~(1.492)	&	0.418	~(0.150)	

%% file: Tab_Result2.tex
Coffee	&	3.446	~(0.500)	&	5.028	~(0.322)	&	4.523	~(0.332)	&	2.070	~(0.441)	\\
Beef	&	2.424	~(0.351)	&	12.885	~(1.579)	&	11.233	~(1.874)	&	1.494	~(0.183)	\\
Blue fish	&	1.892	~(0.106)	&	26.919	~(3.321)	&	19.074	~(1.130)	&	1.461	~(0.171)	\\
Chicken	&	0.792	~(0.363)	&	8.721	~(0.666)	&	4.068	~(0.454)	&	0.938	~(0.088)	\\
Rocks	&	0.829	~(0.101)	&	5.118	~(0.308)	&	3.141	~(0.389)	&	0.965	~(0.125)	\\
Salmon	&	6.575	~(1.287)	&	4.488	~(0.269)	&	5.283	~(0.354)	&	3.772	~(2.572)	\\
Wine	&	0.671	~(0.184)	&	83.064	~(13.03)	&	23.395	~(3.137)	&	0.690	~(0.154)	\\
Cotton\tnote{*1}	&	1.179	~(0.171)	&&&
%	-1095.649	~(7614.631)	&	296.033	~(851.050)	&	
-6.721	~(13.69)	\\
Beer	&	-0.701	~(0.796)	&	17.461	~(1.425)	&	13.092	~(1.011)	&	0.195	~(0.606)	\\
Crabs	&	-1.313	~(0.590)	&	1.498	~(0.061)	&	1.562	~(0.063)	&	1.132	~(0.021)	\\
Shrimps	&	-2.551	~(1.168)	&	2.604	~(0.080)	&	1.698	~(0.055)	&	1.110	~(0.020)	\\
Flat fish	&	-1.796	~(1.116)	&	30.700	~(5.020)	&	1.544	~(0.351)	&	0.287	~(0.493)	

%% file: Tab_Service2.tex
1)	Commercial	&	1.024	~(0.032)	&	0.400	~(0.092)	&	3.697	~(0.736)	&	-0.024	~(0.032)	\\
2)	Construction	&	1.107	~(0.050)	&	0.295	~(0.213)	&	2.981	~(1.348)	&	-0.104	~(0.047)	\\
3)	Financial	&	1.115	~(0.049)	&	0.360	~(0.150)	&	4.009	~(1.121)	&	-0.111	~(0.045)	\\
4)	Goods-related	&	0.794	~(0.031)	&	0.430	~(0.139)	&	4.066	~(0.949)	&	0.226	~(0.036)	\\
5)	Insurance	&	0.840	~(0.035)	&	0.464	~(0.137)	&	4.398	~(0.935)	&	0.172	~(0.041)	\\
6)	Telecom	&	1.066	~(0.054)	&	0.298	~(0.130)	&	3.246	~(0.981)	&	-0.065	~(0.052)	\\
7)	Transport	&	1.006	~(0.027)	&	0.385	~(0.110)	&	3.422	~(0.801)	&	-0.006	~(0.026)	\\
8)	Travel	&	1.063	~(0.039)	&	0.485	~(0.148)	&	3.848	~(0.985)	&	-0.061	~(0.037)

%% file: Tab_IIV2.tex
Coffee	&	5.028	~(0.322)	&	4.523	~(0.332)	&	4877	&	0.102	~(0.750)	&	1602	~(0.000)	\\
Beef	&	12.885	~(1.579)	&	11.233	~(1.874)	&	$6$1	&	0.177	~(0.674)	&	299	~(0.000)	\\
Blue fish	&	26.919	~(3.321)	&	19.074	~(1.130)	&	566	&			&	1024	~(0.000)	\\
Chicken	&	8.721	~(0.666)	&	4.068	~(0.454)	&	221	&	0.009	~(0.923)	&	565	~(0.000)	\\
Rocks	&	5.118	~(0.308)	&	3.141	~(0.389)	&	1705	&	0.252	~(0.616)	&	585	~(0.000)	\\
Salmon	&	4.488	~(0.269)	&	5.283	~(0.354)	&	3321	&	0.117	~(0.732)	&	953	~(0.000)	\\
Wine	&	83.064	~(13.031)	&	23.395	~(3.137)	&	34	&	0.173	~(0.678)	&	955	~(0.000)	\\
 Cotton\tnote{*4}	&&&&& \\ %	-1095.649	~(7614.631)	&	296.033	~(851.050)	&	0	&			&	707	~(0.000)	 \\
Beer	&	17.461	~(1.425)	&	13.092	~(1.011)	&	255	&	3.707	~(0.054)	&	1508	~(0.000)	\\
Crabs	&	1.498	~(0.061)	&	1.562	~(0.063)	&	1088	&			&	112	~(0.000)	\\
Shrimps	&	2.604	~(0.080)	&	1.698	~(0.055)	&	0	&			&	0 ~(0.000)	\\
Flat fish	&	30.700	~(5.020)	&	1.544	~(0.351)	&	544	&	2.566	~(0.109)	&	138 ~(0.000)	

%% file: Tab_HScodes.tex
Beef &\texttt{0201.10-000
0201.20-000
0201.20-010
0201.20-090
0201.30-010
0201.30-020
0201.30-030
0201.30-090
0202.10-000
0202.20-000
0202.20-010
0202.20-090
0202.30-010
0202.30-020
0202.30-030
0202.30-090
}\\ \hline
Beer &\texttt{2203.00-000
}\\ \hline
Blue fish &\texttt{0302.44-000
0302.45-000
0302.61-010
0302.64-000
0303.53-100
0303.54-000
0303.55-000
0303.59-110
0303.59-120
0303.71-010
0303.71-090
0303.74-000
0303.79-021
0303.79-023
0303.79-029
0303.89-121
0303.89-129
}\\ \hline
Chicken &\texttt{0207.11-000
0207.12-000
0207.13-100
0207.13-200
0207.14-100
0207.14-210
0207.14-220
}\\ \hline
Coffee &\texttt{0901.11-000
}\\ \hline
Crabs &\texttt{0306.11-200
0306.12-200
0306.14-010
0306.14-020
0306.14-030
0306.14-040
0306.14-090
0306.24-110
0306.24-121
0306.24-129
0306.24-130
0306.24-140
0306.24-150
0306.24-190
0306.24-200
0306.26-300
0306.33-110
0306.33-121
0306.33-129
0306.33-130
0306.33-140
0306.33-150
0306.33-190
0306.93-900
}\\ \hline
Flat fish &\texttt{0301.99-220
0302.21-000
0302.23-000
0302.24-000
0302.29-000
0303.31-000
0303.32-000
0303.33-000
0303.34-000
0303.39-000
0304.43-000
0304.83-000
}\\ \hline
Salmon &\texttt{0302.11-000
0302.12-011
0302.12-019
0302.12-020
0302.13-011
0302.13-012
0302.13-019
0302.14-000
0302.19-000
0303.11-000
0303.12-010
0303.12-090
0303.13-000
0303.14-000
0303.19-000
0303.19-010
0303.19-090
0303.21-000
0303.22-000
0304.29-950
0304.41-000
0304.42-000
0304.81-000
0304.82-000
0304.89-220
}\\ \hline
Shrimps &\texttt{0306.13-000
0306.16-000
0306.16-200
0306.17-000
0306.17-100
0306.17-200
0306.19-010
0306.19-090
0306.19-100
0306.19-190
0306.19-200
0306.19-290
0306.23-111
0306.23-119
0306.23-190
0306.23-200
0306.27-111
0306.27-119
0306.27-190
0306.27-200
0306.27-300
0306.29-110
0306.29-120
0306.29-190
0306.29-210
0306.29-220
0306.29-290
0306.36-110
0306.36-190
0306.36-900
0306.39-100
0306.39-900
0306.95-100
0306.95-900
0306.99-910
0306.99-990
}\\ \hline
Rocks &\texttt{6801.00-000
}\\ \hline
Wine &\texttt{2204.10-000
2204.21-010
2204.21-020
2204.22-000
2204.29-010
2204.29-090
2204.30-111
2204.30-119
2204.30-191
2204.30-200
} \\ \hline

%% file: Tab_STRI.tex
2)	&	1.229	~(0.200)	&	0.077	~(0.412)	&	-0.136	~(0.299)	&	-0.237	~(0.213)	&	0.080	~(0.427)	\\
3)	&	1.148	~(0.204)	&	-0.760	~(0.291)	&	0.436	~(0.200)	&	-0.139	~(0.181)	&	-0.714	~(0.274)	\\
5)	&	1.394	~(0.332)	&	-0.078	~(0.338)	&	0.587	~(0.155)	&	-0.320	~(0.218)	&	-0.063	~(0.275)	\\
6)	&	1.831	~(0.377)	&	-0.581	~(0.886)	&	0.328	~(0.167)	&	-0.653	~(0.225)	&	-0.456	~(0.713)	\\
7)	&	1.019	~(0.128)	&	-0.815	~(0.604)	&	0.535	~(0.288)	&	-0.018	~(0.125)	&	-0.807	~(0.612)	